\documentclass{article}

\usepackage{PRIMEarxiv}

\usepackage[utf8]{inputenc} 
\usepackage[T1]{fontenc}    
\usepackage{hyperref}       
\usepackage{url}            
\usepackage{booktabs}       
\usepackage{amsfonts}       
\usepackage{nicefrac}       
\usepackage{microtype}      
\usepackage{lipsum}
\usepackage{fancyhdr}       
\usepackage{graphicx}       
\usepackage{amsmath}
\usepackage{biblatex}
\usepackage{hyperref}
\graphicspath{{media/}}     
\title{Virtual Avatar Stream: A cost-down approach to the metaverse experience

}

\author{
  Joseph Chang \\
  \texttt{josephjbc@gmail.com} \\
}

\begin{document}
\maketitle

\begin{abstract}
The Metaverse through VR headsets is a rapidly growing concept, but the high cost of entry currently limits access for many users. This project aims to provide an accessible entry point to the immersive Metaverse experience by leveraging web technologies. The platform developed allows users to engage with rendered avatars using only a web browser, microphone, and webcam. By employing the WebGL and MediaPipe face tracking AI model from Google, the application generates real-time 3D face meshes for users. It uses a client-to-client streaming cluster to establish a connection, and clients negotiate SRTP protocol through WebRTC for direct data streaming. Additionally, the project addresses backend challenges through an architecture that is serverless, distributive, auto-scaling, highly resilient, and secure. The platform offers a scalable, hardware-free solution for users to experience a near-immersive Metaverse, with the potential for future integration with game server clusters. This project provides an important step toward a more inclusive Metaverse accessible to a wider audience.
\end{abstract}

\keywords{Metaverse \and Face Mesh \and Streaming}

\section{Introduction}
The Metaverse is a digital universe that seeks to replicate the physical world through technologies \cite{metaverse}. With the advancement of virtual reality (VR) technologies, VR headsets have become a primary interface for accessing the Metaverse. The immersive experience provided by VR headsets has the potential to revolutionize human interactions. These devices can capture upper body movement data from the user and create a virtual avatar, allowing users to move in a virtual digital world by moving their physical bodies. In addition, VR headsets have two screens and speakers. The screens produce parallax based on the distance of objects in the virtual world, mimicking the sense of 3D vision humans have in the real world. The speakers produce stereophonic sound to simulate the 3D vision and sound direction of the real world. By combining body movement data with the digital interface, VR headsets create a virtual reality experience that simulates the spatial environment of the physical world. This technology enables users to interact with one another in a virtual yet spatial environment, regardless of their physical location. This has numerous applications, including virtual conferences, gaming, and education.

However, the high cost of entry has hindered consumers from investing in this technology, particularly those who do not understand or have not experienced the potential of the immersive Metaverse. Here, we propose to provide an entry-level immersive Metaverse experience by leveraging face tracking, 3D mesh rendering, and live streaming technologies. Furthermore, our solution leads to other applications, such as a virtual Omegle chat and a setup for virtual YouTubers. The former provides a safe space for one-on-one video chats with random users while displaying facial movements and expressions on avatar mesh. The latter eliminates the need for costly tracking headsets by using software to track and animate characters with a webcam.

To offer a partially immersive Metaverse experience, we provide a solution to replace the 3D conference call feature from Meta, which requires a costly VR headset \cite{meta_quest_price}. In brief, we use an AI Face Mesh model to track the user's facial expressions, position, and orientation, to render 3D mesh data into a 2D screen through a rendering pipeline done on WebGL, allowing for a simulated 3D conference experience using only a webcam, microphone, and web browser. The current implementation enables two users to engage in a virtual video call while displaying their avatar faces. 

In summary, we demonstrate that companies like Meta can use this product to offer an entry-level Metaverse experience through a web portal rather than requiring consumers to invest in expensive hardware. By providing these accessible applications, our solution is expected to bridge the gap for consumers to experience the immersive Metaverse.

\label{sec:headings}
\section{Face Tracking}
\begin{figure}
    \begin{minipage}[t][6.7cm][t]{.4\linewidth}
        \includegraphics[height=4cm]{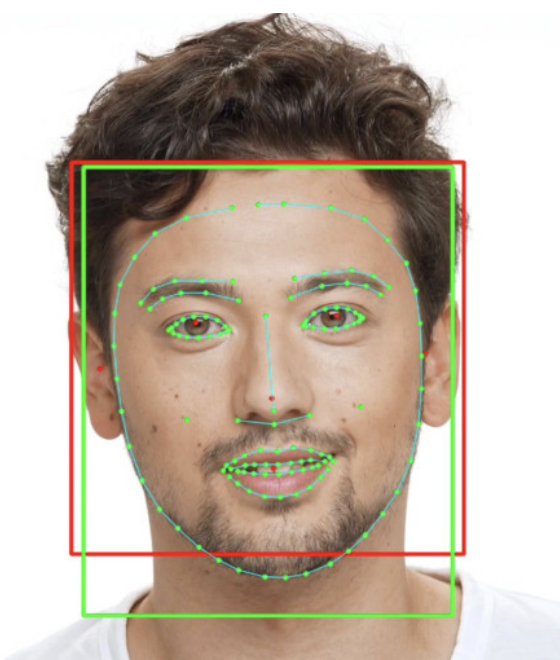}
        \centering
        \caption{Face Detection Model output example adapted from BlazeFace \cite{faceDetection}. The green line indicates the output rectangle used in the Face Mesh pipeline.}
        \label{fig:faceDetection}
    \end{minipage}%
    \hspace{1.5cm}
    \begin{minipage}[t][6.7cm][t]{.5\linewidth}
        \includegraphics[height=4cm]{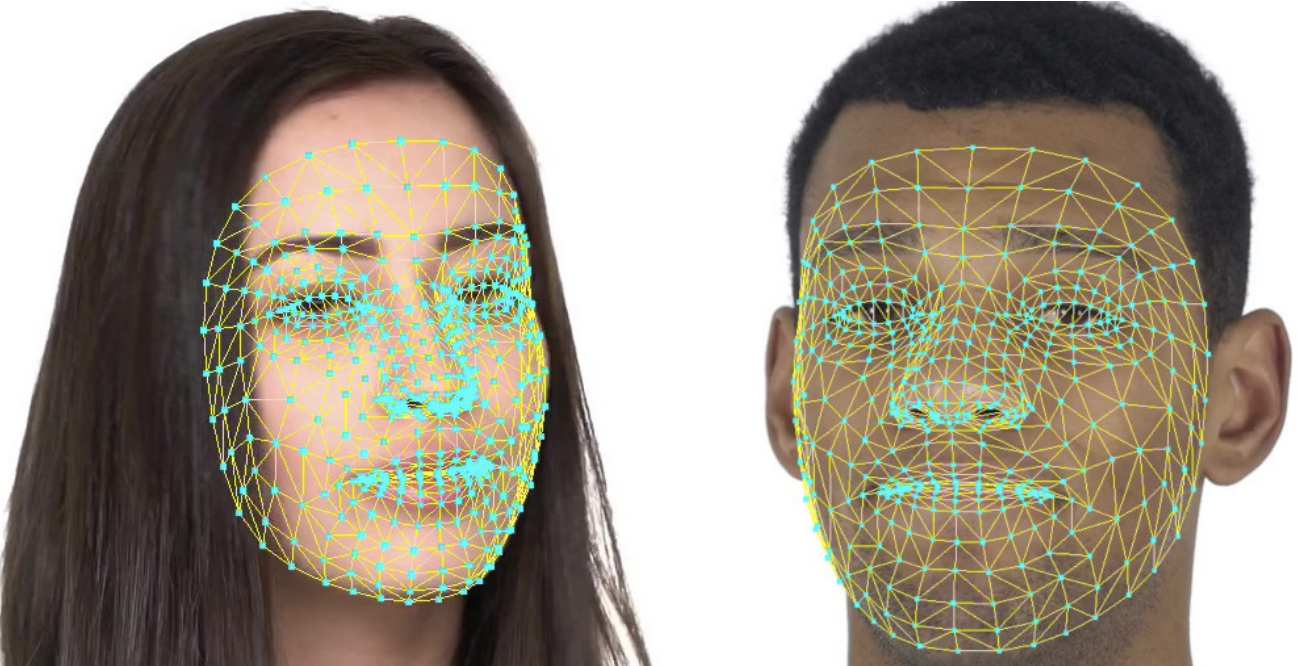}
        \centering
        \caption{Face Landmark Model output examples adapted from \cite{faceLandmarks}.}
        \label{fig:faceLandmarks}
    \end{minipage}
\end{figure}

This product utilizes the Google MediaPipe library's AI model, specifically, the MediaPipe Face Mesh model. The model takes images as input and generates 468 feature coordinates for each frame on the user’s face as output \cite{doc_mediapipe_facemesh_model}. The model is composed of two deep neural network sub-models: the Face Detection Model and the Face Landmark Model. The Face Detection Model employs the BlazeFace model \cite{faceDetection} to locate, crop, and predict rotations for the faces in a video. The cropped face with rotation is then transformed into a flat upright rectangle of 128 or 256 square pixels, which is fed into the Face Landmark Model to generate 3D landmark vectors \cite{faceLandmarks}. These vectors can be used to calculate the predicted 3D facial landmark coordinates and mapped back to the cropped face with rotation. The Face Detection Model uses a CNN architecture \cite{faceDetection}, while the Face Landmark Model employs ResNet, which is also a CNN architecture \cite{faceLandmarks}. Both models have been optimized for GPU acceleration to enable applications on personal devices \cite{faceDetection, faceLandmarks}. Examples of the models' outputs can be seen in figures \ref{fig:faceDetection} and \ref{fig:faceLandmarks}. The system captures real-time images from the user's webcam and processes them with the Face Mesh model to retrieve facial landmarks, which are then used for rendering.

\section{Rendering}
\subsection{Graphic Pipeline}

The platform uses WebGL for graphic rendering. WebGL is a JavaScript implementation of OpenGL, a widely-used graphics API that enables developers to construct GPU-accelerated rendering pipelines. These pipelines consist of several stages, including transformation, vertex shading, fragment shading, and perspective projection. The transformation stage involves a series of operations that are represented by equations \ref{eqn:homogeneous} to \ref{eqn:transform}. Homogeneous coordinates, a 4x4 matrix representation for 3D computation, are used in all 3D operations to facilitate translation.

\begin{align}
&Homogeneous \ coordinates \quad & P_{init} &= \begin{bmatrix} x \\ y \\ z \\ 1 \end{bmatrix}
\label{eqn:homogeneous}
\\
&Scale \quad & S &= \begin{bmatrix} 
S_x & 0 & 0 & 0 \\
0 & S_y & 0 & 0 \\
0 & 0 & S_z & 0 \\
0 & 0 & 0 & 1 
\end{bmatrix}
\label{eqn:scale}
\\
&Rotation \quad & R_x (\theta) &= \begin{bmatrix} 
1 & 0 & 0 & 0 \\
0 & cos \theta & -sin \theta & 0 \\
0 & sin \theta & cos \theta & 0 \\
0 & 0 & 0 & 1 
\end{bmatrix}
\label{eqn:rotate}
\\
&Translation \quad & T &= \begin{bmatrix} 
1 & 0 & 0 & T_x \\
0 & 1 & 0 & T_y \\
0 & 0 & 1 & T_z \\
0 & 0 & 0 & 1 
\end{bmatrix}
\label{eqn:translate}
\\
&Transformation \quad & P_{after} &= M_{trans} \times P_{init}
\label{eqn:transform}
\end{align}

However, homogeneous coordinates are not required for perspective projection, and the projection matrix can be reduced to a 3x3 matrix to reduce computational overhead. Once a camera and projection screen are defined, the perspective projection algorithm can be implemented using equations \ref{eqn:perspective} and \ref{eqn:projection}. This approach allows for more efficient rendering of 3D scenes.

\begin{align}
\label{eqn:perspective}
\begin{split}
\begin{bmatrix} 
f_x \\
f_y \\
f_w 
\end{bmatrix} 
& = 
\begin{aligned}[t]
\begin{bmatrix} 
1 && 0 && \frac{e_x}{e_z} \\
0 && 1 && \frac{e_y}{e_z} \\
0 && 0 && \frac{1}{e_z} 
\end{bmatrix} 
\begin{bmatrix} 
1 && 0 && 0 \\
0 && cos \theta_x && sin \theta_x \\
0 && -sin \theta_x && cos \theta_x 
\end{bmatrix} 
\begin{bmatrix} 
cos \theta_y && 0 && -sin \theta_y \\
0 && 1 && 0 \\
sin \theta_y && 0 && cos \theta_y 
\end{bmatrix}
\\ 
\begin{bmatrix} 
cos \theta_z && sin \theta_z && 0 \\
-sin \theta_z && cos \theta_z && 0 \\
0 && 0 && 1 
\end{bmatrix}
\begin{pmatrix}
\begin{bmatrix} 
p_x \\
p_y \\
p_z 
\end{bmatrix} 
- \begin{bmatrix} 
c_x \\
c_y \\
c_z 
\end{bmatrix}
\end{pmatrix}
\end{aligned}
\end{split}
\\
\begin{bmatrix}
b_x \\ b_y
\end{bmatrix} & = 
\begin{bmatrix}
f_x / f_w \\ f_y / f_w
\end{bmatrix}
\label{eqn:projection}
\end{align}

\begin{align*}
&p: \ The \ coordinates \ of \ a \ vertex
\\
&c: \ The \ coordinates \ of \ the \ camera
\\
&e: \ The \ display \ surface \ of \  the \ projection \ screen
\\
&f: Homogenous \ coordinates \ of \ 2D \ point
\\
&b: 2D \ display \ coordinates
\end{align*}

\subsection{Vertices and Calibration}

\begin{figure}
    \includegraphics[width=10cm]{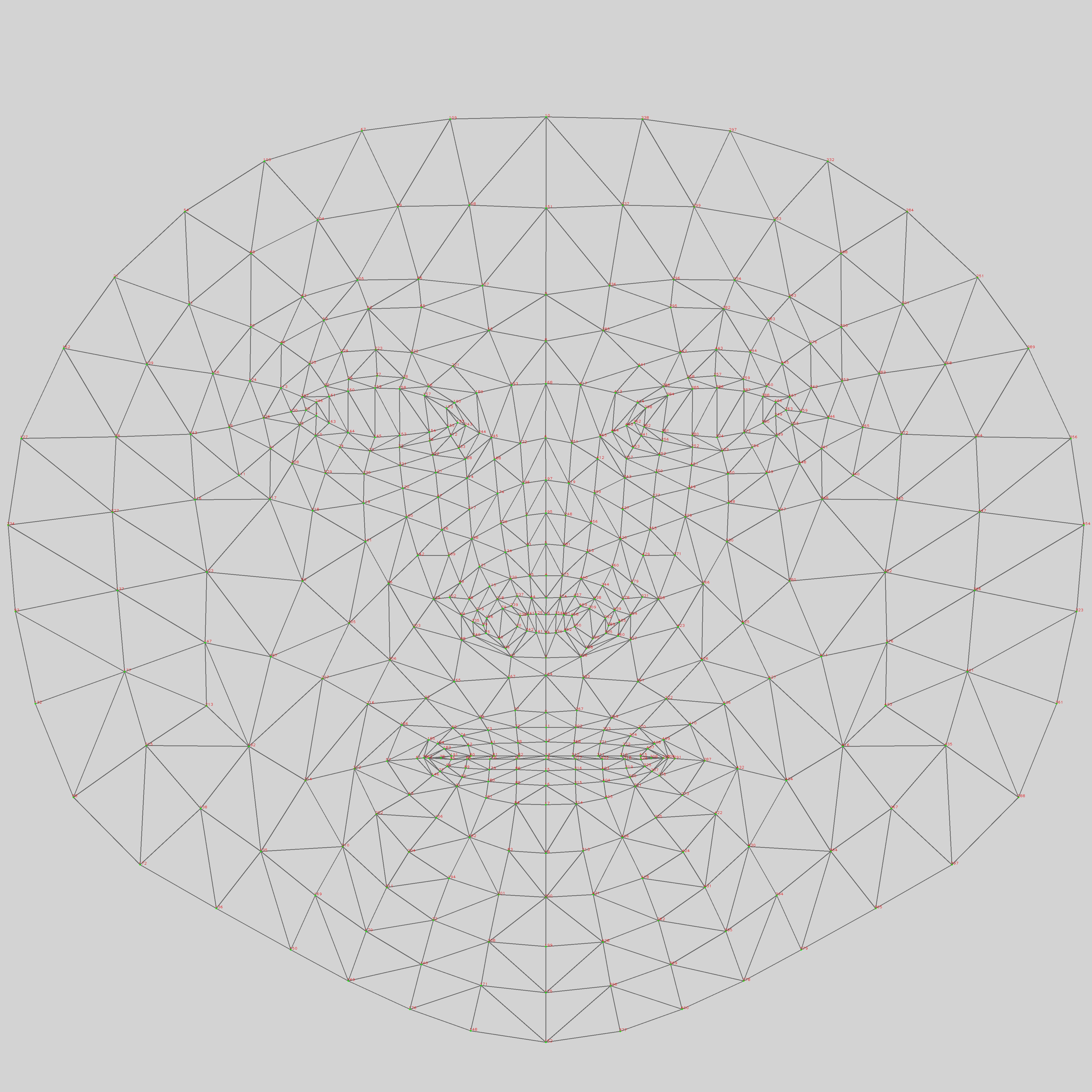}
    \centering
    \caption{Face mesh with landmark number labels adapted from \cite{pic_mediapipe_mesh}.}
    \label{fig:facemesh}
\end{figure}

The vertices of the face mesh are computed by grouping the landmarks into triangles, as shown in figure \ref{fig:facemesh}. The vertices are generated by performing both clockwise and counterclockwise rotations, resulting in a mesh that is filled from both sides. As a result, if the mesh is rotated by 180 degrees, it will not appear transparent.

To ensure the user's face is calibrated with the 3D-rendered avatar face, spatial translation and rotation are applied. It expects the user to have their face-centered and upright at the initial frame. And it transforms the avatar's face so that the initial frame's nose is centered, and the triangle formed by the nose and two eyelashes lies on a flat surface in the graphic space with its normal pointing outward from the screen. The calibration conditions are expressed in equations \ref{eqn:center_nose}, \ref{eqn:up_vec}, \ref{eqn:right_vec}, and \ref{eqn:orientation_out}. The facial landmarks used to construct the face mesh are depicted in figure \ref{fig:facemesh}.

\begin{align}
landmark_{5} &\Leftarrow \begin{bmatrix} 0 \\ 0  \\ 0 \end{bmatrix}
\label{eqn:center_nose}
\\
\vec{up} &= (\frac{landmark_{52}-landmark_{282}}{2})-landmark_{1}
\label{eqn:up_vec}
\\
\vec{right} &= landmark_{282} - landmark_{52}
\label{eqn:right_vec}
\\
\hat{up} \times \hat{right} &\Leftarrow 
\begin{bmatrix} 0 \\ 0  \\ 1 \end{bmatrix} = \hat{z}
\label{eqn:orientation_out}
\end{align}

\begin{align*}
landmark_{5}&: \quad center \ nose \\
    landmark_{1}&: \quad lower \ nose \\
    landmark_{52}&: \quad left \ eyelash \\
    landmark_{282}&: \quad right \ eyelash
\end{align*}

To achieve the condition given in equation \ref{eqn:orientation_out}, the calibration algorithm measures the off-angles of the up and right vectors and rotates the face mesh by these off-angles using quaternion angles. The calibration algorithm performs equations \ref{eqn:quaternion_up}, \ref{eqn:quaternion_right}, and \ref{eqn:quaternion_facemesh} to ensure the orientation condition is met.

\begin{align}
Q_{up} &= QuaternionAngle(\hat{up}, \hat{y})
\label{eqn:quaternion_up}
\\
Q_{right} &= QuaternionAngle(\hat{right}, \hat{x})
\label{eqn:quaternion_right}
\\
Q_{FaceMeshRotation} &= Q_{up} \times Q_{right}
\label{eqn:quaternion_facemesh}
\end{align}

\subsection{Quaternions}

Quaternions are a four-dimensional system that uses $\vec{x}$, $\vec{y}$, $\vec{z}$, and $\vec{w}$ to represent a rotation. An alternative way to represent a rotation is to use Euler angles of $\phi$, $\theta$, and $\psi$ with a rotation order such as XYZ to achieve $roll$, $yaw$, and $pitch$. The quaternion system is used because it avoids Gimbal lock. The matrix representation of Euler rotation in XYZ rotating sequence and the derivation of quaternions from the Euler angles are shown in equations \ref{eqn:euler_rotation} and \ref{eqn:quaternions}, adapted from \cite{quaternions}.

\begin{align}
\label{eqn:euler_rotation}
\begin{split}
\begin{bmatrix} x_{after} \\ y_{after} \\ z_{after} \end{bmatrix}
&
\begin{aligned}[t]
&= R_z(\psi)R_y(\theta)R_x(\phi)
\begin{bmatrix} x_{init} \\ y_{init} \\ z_{init} \end{bmatrix}
\\
&= \begin{bmatrix}
cos\psi & -sin\psi & 0 \\
sin\psi & cos\psi & 0 \\
0 & 0 & 1
\end{bmatrix}
\begin{bmatrix}
cos\theta & 0 & sin\theta \\
0 & 1 & 0 \\
-sin\theta & 0 & cos\theta
\end{bmatrix}
\begin{bmatrix}
1 & 0 & 0 \\
0 & cos\phi & -sin\phi \\
0 & sin\phi & cos\phi
\end{bmatrix}
\begin{bmatrix} x_{init} \\ y_{init} \\ z_{init} \end{bmatrix}
\end{aligned}
\end{split}
\\
\label{eqn:quaternions}
\begin{bmatrix} Q_x \\ Q_y \\ Q_z \\ Q_w \end{bmatrix} &= 
\begin{bmatrix} 
sin\frac{\phi}{2}cos\frac{\theta}{2}cos\frac{\psi}{2} -
cos\frac{\phi}{2}sin\frac{\theta}{2}sin\frac{\psi}{2}
\\
cos\frac{\phi}{2}sin\frac{\theta}{2}cos\frac{\psi}{2} +
sin\frac{\phi}{2}cos\frac{\theta}{2}sin\frac{\psi}{2}
\\
cos\frac{\phi}{2}cos\frac{\theta}{2}sin\frac{\psi}{2} - 
sin\frac{\phi}{2}sin\frac{\theta}{2}cos\frac{\psi}{2} 
\\
cos\frac{\phi}{2}cos\frac{\theta}{2}cos\frac{\psi}{2} +
sin\frac{\phi}{2}sin\frac{\theta}{2}sin\frac{\psi}{2}
\end{bmatrix}
\end{align}

\section{Stream}
\subsection{Infrastructure}

The platform is deployed using AWS (Amazon Web Services), and the architecture described is a best practice cluster deployment suggested by AWS \cite{doc_aws_ecs}. The resources for the application are grouped within a VPC (Virtual Private Cloud) consisting of public and private subnets. The ALB (Application Load Balancer) is deployed within the public subnets to receive the traffic from the public network and distribute them to the cluster. The ECS (Elastic Container Service) cluster and Aurora database are deployed within the private subnets to that both of them are not directly accessible from the public network. The redundant subnets ensure the application survives on a zone outage.

\begin{figure}
    \includegraphics[width=\linewidth]{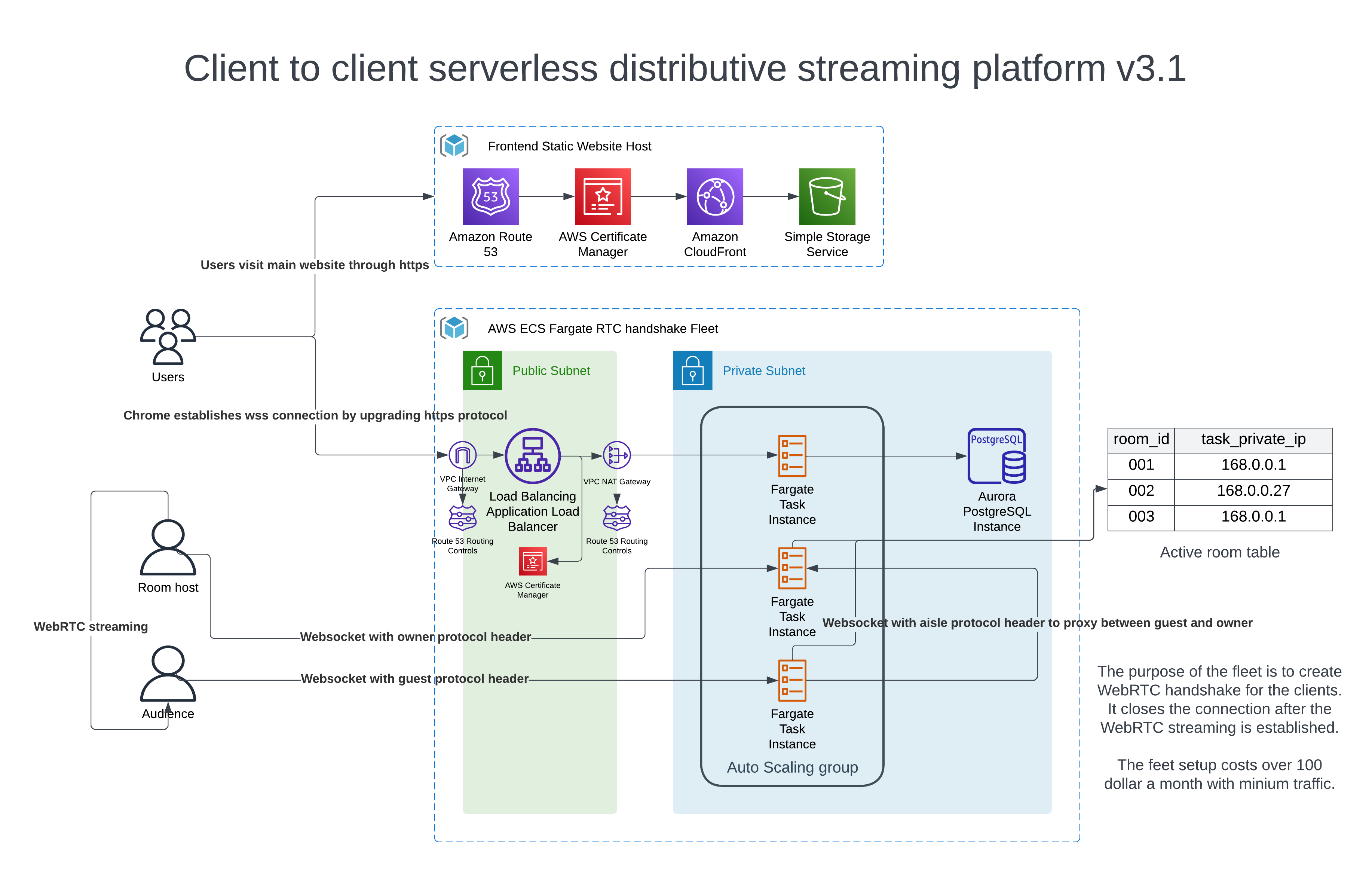}
    \centering
    \caption{architecture diagram of virtual avatar stream}
    \label{fig:architecture}
\end{figure}

The system is distributive and horizontally scalable, with multiple instances monitored for CPU and memory usage to achieve resiliency and auto-scaling. Terraform, an infrastructure-as-code software tool \cite{doc_terraform}, is used to manage the deployment procedure, which allows for robust infrastructure management. This architecture is highly versatile and can be used to support a wide range of applications, including popular services like Discord or online games that need to handle millions of users concurrently. The architecture diagram is shown in figure \ref{fig:architecture}.

\subsection{Protocols}

In this system, WebSocket, a TCP data streaming protocol, is used to establish and maintain the connection between the front end and the backend. The streaming protocol is essential in helping front-end clients find each other and negotiate peer-to-peer protocols. Since negotiation requires back-and-forth data exchange until an agreement is reached, streaming becomes necessary. As the system is distributed, clients with the same unique room ID may not always land on the same cluster instance. In the event that clients open WebSocket on different distributed cluster instances, the instance will look up the private IP in the database and create a proxy connection between instances to enable data swapping.

WebRTC, an SRTP protocol that streams data from client to client directly \cite{doc_webrtc}, is used to avoid the backend handling sensitive data such as user voice and face mesh. The front end uses the backend to negotiate the offer, answer, and ICE candidate for WebRTC, and the client starts to send audio and face mesh data, including the mesh translation and rotation, to another client once the peer-to-peer connection is established using WebRTC. The voice is sent through a streaming track, and the face mesh is sent through a data channel.

\subsection{Compression}
When the internet speed is slow or the distance between clients is long, the data channel can suffer from network congestion, which is solved by compressing the landmark data to float 16, the half-precision floating-point. The payload size is compressed to 2838 bytes per frame, and the max fps is set to 30, so the data rate is about 85 KB/s. The data rate of a typical YouTube video, 1080p 30fps with an H264 compression ratio of 2000:1 \cite{compression_ratio}, is 93 KB/s which shows the platform data rate of 85 KB/s is operable. The derivation of the H264 video data rate is shown in equation \ref{eqn:video_data}.

\begin{equation}
1080 \times 1920 \times 3 \frac{bytes}{pixel} \times 30 fps \times \frac{1}{2000} = 93 KB/s
\label{eqn:video_data}
\end{equation}

\begin{figure}
    \includegraphics[width=10cm]{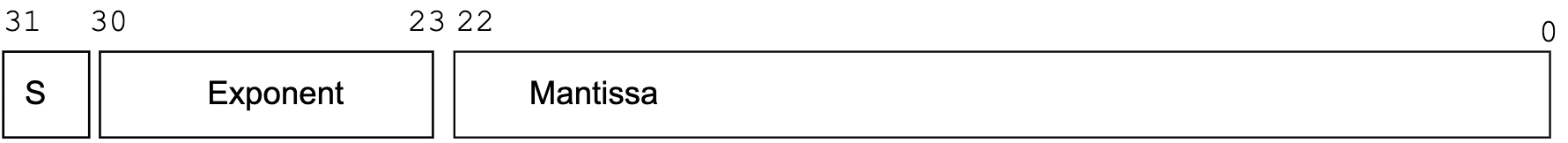}
    \centering
    \caption{IEEE single-precision floating-point (float 32) standards adapted from \cite{doc_float32}.}
    \label{fig:float32}

    \includegraphics[width=10cm]{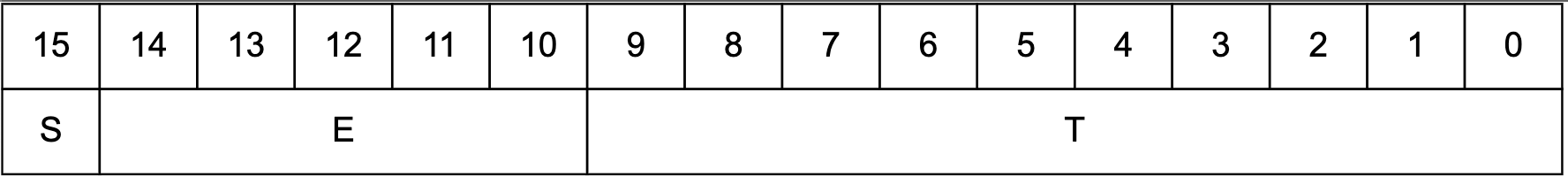}
    \centering
    \caption{IEEE half-precision floating-point (float 16) standards adapted from \cite{doc_float16}. The "T" here is mantissa.}
    \label{fig:float16}
\end{figure}

The IEEE standards for float 32 and float 16 are depicted in figures \ref{fig:float32} and \ref{fig:float16}, respectively. To convert a float 32 to float 16, the sign bit remains unchanged, the mantissa is truncated, and the leftmost 10 bits from float 32 are stored in the mantissa of float 16. Meanwhile, the exponent undergoes the arithmetic in equation \ref{eqn:exponent}, but special conditions of overflow and underflow must be taken into account.

\begin{equation}
E_{float16} = E_{float32} - \text{0x7F} + \text{0xF} = E_{float32} - 127 + 15
\label{eqn:exponent}
\end{equation}

\section{Results}
The project successfully demonstrates the potential of delivering a VR headset Metaverse experience on the web using conventional hardware that most customers already have. It utilizes a combination of open-source libraries and web technologies, including Google MediaPipe, ThreeJS, WebSocket, WebRTC, and AWS. The final product is available on the website: \href{https://virtualavatar.trip1elift.com/}{virtualavatar.trip1elift.com}. The source code is on the repository: \href{https://github.com/Trip1eLift/virtual-avatar}{github.com/Trip1eLift/virtual-avatar}, with a user guide: \href{https://github.com/Trip1eLift/virtual-avatar/blob/main/userguide/README.md}{github.com/Trip1eLift/virtual-avatar/tree/main/userguide}.

Upon the first visit, the user is prompted to grant browser access to the camera and microphone. The camera is used to capture the image of the user's face, which is processed by the 3D face tracking AI model of MediaPipe to obtain facial landmarks. These landmarks are rendered into 2D graphics using the ThreeJS library, with randomly generated colors. The microphone is utilized when the user engages in a room chat with another user to collect audio data. The website aligns the user's nose and eyelashes with a flat surface of a normal vector pointing out of the screen in the space. The user can modify, store, and load calibration settings manually. Figure \ref{fig:calibration} shows the calibration interface.https://www.overleaf.com/project/6408de81f53a680b6c52b908

\begin{figure}
    \includegraphics[width=10cm]{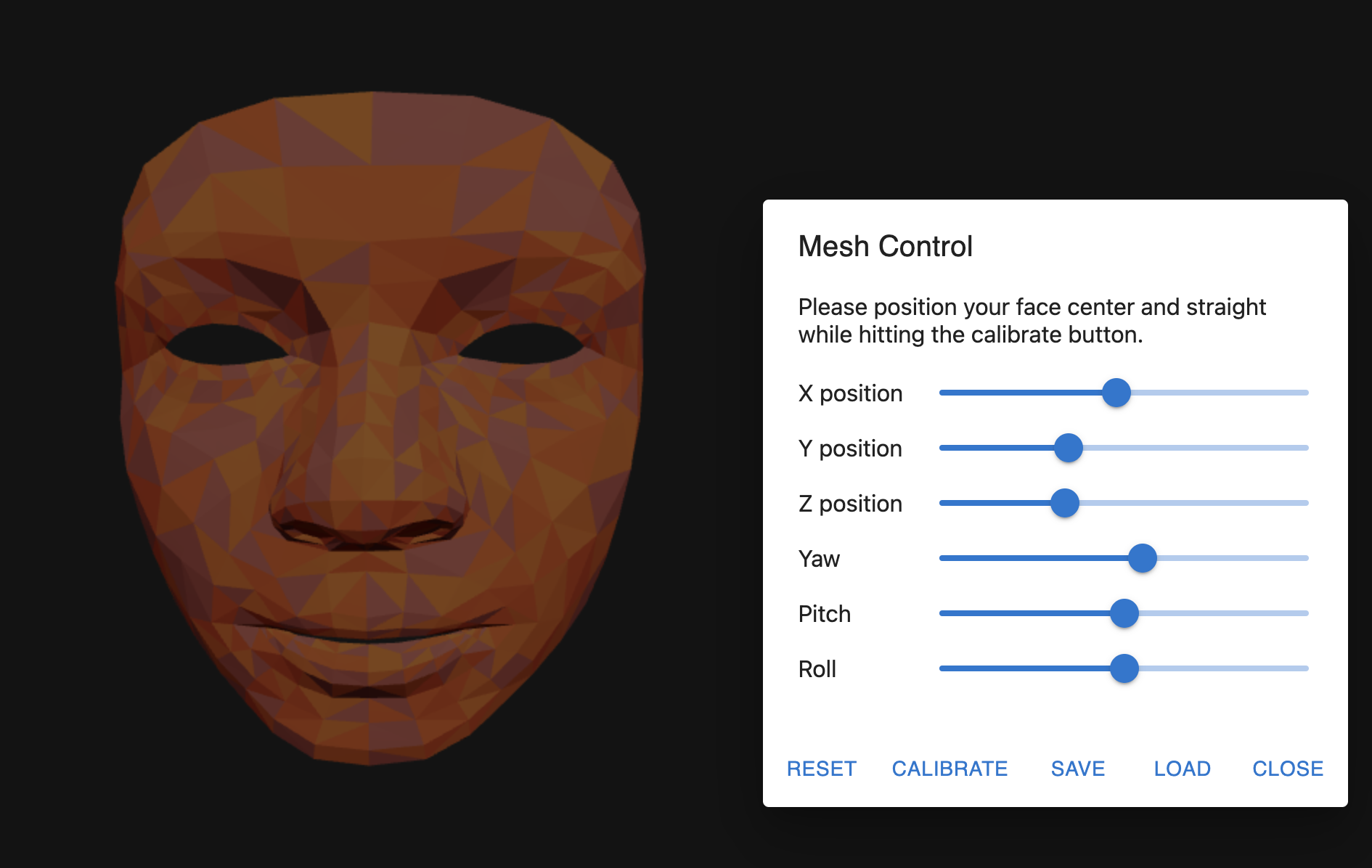}
    \centering
    \caption{virtual avatar stream mesh control panel for calibration}
    \label{fig:calibration}
\end{figure}

Once calibration settings are configured, the user can begin streaming. Two types of users are supported: room owners and guests. The room owner can create a room by clicking the "create room" button in the streaming control panel. The streaming cluster generates a unique room ID that the room owner must share with the guest through an external channel. Since the website doesn't have a user account system, it doesn't support friend calling. The room owner maintains a WebSocket connection to the backend and waits for the guest to join. During scale-down, both the room owner and guest may land on the same distributive cluster instance, allowing the front ends to negotiate the WebRTC protocol on the same instance. During traffic scaling, the room owner and guest may land on different distributive cluster instances. The instance the guest is on uses the room ID to obtain the private IP of the instance the room owner is on from the database, and the guest instance proxies WebSocket data to the owner's instance. The two parties negotiate the WebRTC protocol through the proxy.

\begin{figure}
    \includegraphics[width=10cm]{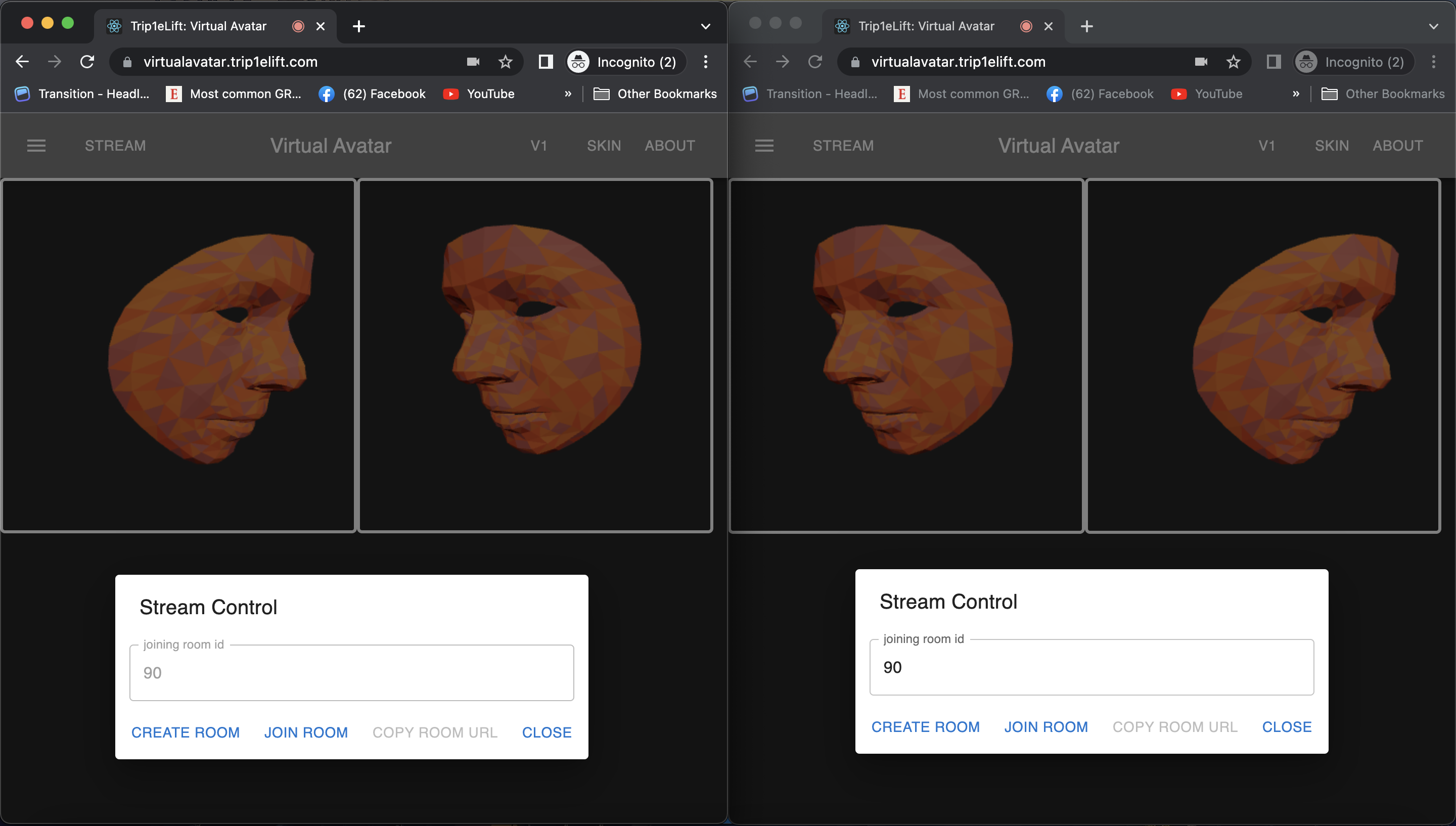}
    \centering
    \caption{virtual video call between two users in room 90 using virtual avatar stream}
    \label{fig:finalProduct}
\end{figure}

After the WebRTC connection between the front ends is established, both hang up their WebSocket connections to the backend cluster. Mesh data streaming and audio data streaming commence. The mesh data is compressed to 43 KB/s to prevent network congestion. Figure \ref{fig:finalProduct} illustrates the streaming interface. This product demonstrates that an immersive Metaverse experience with VR headsets can be brought to the web without requiring a VR headset.

The front end of the product is developed using ReactJS \cite{doc_react} and is supplemented with various API documentation resources \cite{doc_mui, doc_mediapipe, doc_webgl, doc_three, doc_react_three, doc_webrtc}. Additionally, the front end is bolstered with the use of npm packages \cite{npm_mui, npm_mediapipe, npm_three, npm_react_three, npm_react_webcam, npm_react_draggable}. The back-end is built with golang \cite{doc_golang} and leverages open-source packages \cite{go_websocket, go_pgx}. The system is containerized using Docker \cite{doc_docker} and is deployed to AWS \cite{doc_aws_ecs} with the help of Terraform \cite{doc_terraform}. The product employs Postgres \cite{doc_postgres} as its database system.

\section{Discussion}
The current product demonstrates that it is possible to replace the facial movements and expressions of a VR headset on a web-based platform using conventional hardware. However, the potential of this product is much higher than what it currently offers. Many VR headsets on the market come with VR controllers that allow users to control their hand movements in the virtual world. While Google MediaPipe supports upper body spine, arms, and finger tracking, this platform does not support these body parts due to the limitations of its basic graphic engine. To render the details of arm and finger movements, a more advanced physics engine is needed to avoid mesh clipping or other graphic glitches.

Another area that can be improved upon is face mesh rendering. The Metaverse allows users to become whatever they want, like a humanoid with a face of a cat or a dog. The challenge of implementing such a feature lies in the graphic engine. A more advanced physics/graphic engine is needed to support vertices mapping with weight to map the face mesh onto a dog face while interpolating the facial expression.

Additionally, there is a gap between this platform and the Metaverse that runs on a VR headset in terms of users' visual perception due to the output interface. In a VR headset, users can see things in 3D because the two lenses on the VR headset create parallax to mimic 3D visualization. This project uses a traditional computer screen as the output interface, and while the data was captured in 3D, it projects the 3D data into a 2D space to show it to the user. This limitation can be solved by using a red-blue 3D image rendering, which any colored 2D monitor can output. The cost of red-blue 3D glasses made out of cardboard is minimal, so with an upgrade of the graphic engine and a minimal cost, users can experience 3D visuals that they could only experience on a VR headset.

Another limitation of the current platform is that it only supports one-on-one video chat and does not offer 3D spatial sound, which is a standard feature of most Metaverse platforms. However, in the future, this product can be scaled to support a conference room of many people. Implementing 3D spatial sound can be achieved by offsetting the sound of left and right sound tracks and decaying the intensity of the sound based on the energy decay of the inverse-square law.

\section{Conclusion}
In this project, we have demonstrated the potential of creating an immersive Metaverse experience with the usage of conventional cameras and microphones. We have built a scalable client-to-client streaming backend that can handle millions of users, which can be extended to other applications. Our product leverages Google MediaPipe's 3D face tracking model to convert webcam data into face mesh landmarks, which are rendered using ThreeJS into 3D computer graphics. We use WebSocket and WebRTC protocols to enable direct streaming of the face mesh and audio data between two users for a virtual avatar video call. Our compression techniques ensure that network congestion is prevented and the streaming quality is maintained at an acceptable level.

Although the current experience provided by our product is not as immersive as that provided by a VR headset, it offers an entry-level experience for beginners to understand the concept of the Metaverse. With further development, it is possible to enhance the product to provide an experience that is nearly as good as that of a VR headset. We believe that our platform can serve as a stepping stone for future innovations in the field of the Metaverse.

\printbibliography

\end{document}